# Ultra-Fast Device-Free Visible Light Sensing and Localization via Reflection-Based ΔRSS and Deep Learning

Helena Serpi
*Department of Informatics and Telecommunications*
*University of Peloponnese*
Tripoli, Greece
e.serpi@go.uop.gr

Christina (Tanya) Politi
*Department of Electrical and Computer Engineering*
*University of Peloponnese*
Patras, Greece
email tpoliti@uop.gr

*Abstract*— We propose an Ultra-Fast, Device-Free Visible Light Sensing and Positioning system that captures spatiotemporal variations in single-LED VLC channel responses, using ceiling-mounted photodetectors, to accurately and non-intrusively infer human presence and position through optical signal reflection modeling. The system is highly adaptive and ready to serve different real-world sensing and positioning scenarios using one or more ML based models from the library of multi-architecture deep neural network ensembles we have developed.

## I. INTRODUCTION

The adoption of advanced technologies that extend beyond conventional communication methods is set to transform the utilization of mobile networks. Central to this transformation are technologies such as Integrated Communication, Computation, and Caching, Integrated Sensing and Communication (ISAC), and Wireless Energy Transfer [1]. Notably, ISAC is considered revolutionary for Sixth-Generation Wireless Networks (6G) as it merges radio sensing with communication, enabling mobile networks to perceive their surroundings and offer services that surpass basic communication with emphasis to outdoor environments. In recent years, growing interest from both academia and industry in optical wireless communication technologies—exploiting visible and infrared spectrum for last-meter and last-mile applications [2]—has driven significant advancements in visible light communication (VLC), which in turn have propelled the development of visible light sensing (VLS) systems. Like other wireless sensing technologies, these systems are categorized into two types: device-based (active) and device-free (passive), depending on whether the targets, such as individuals or objects of interest, are equipped with an optical receiver like a photodiode or camera [3]. By utilizing the existing LED lighting infrastructure as transmitters in indoor environments, device-free VLS systems can address the growing spectrum abundant, energy efficiency demands for pervasive sensing and communication in indoor environments.

Supported by the continuous progress in artificial intelligence (AI), deep learning-based visible light positioning (DL VLP) and sensing (DL VLS) have emerged as significant research areas. Despite the advancements, most current systems lack simplicity, high accuracy, and real-time operation capabilities [4]. Our aim is to study the design aspects for a system that address these shortcomings, combining simplicity with accuracy and real-time functionality. This work takes advantage of fast ML based calculations of Received Signal Strength (RSS) in a room and introduces a Deep Neural Network (DNN) ensemble model [5, 6] designed for device-free indoor positioning and sensing using VLC (DNN-VLP ensemble). The DNN-VLP system detects human presence by observing RSS variations as received by a set of ceiling receivers, caused by reflections when a person enters the indoor area. Photodetectors (PDs) installed on a ceiling capture these fluctuations, and the DNN-VLP ensemble model (that may run on a single board computer like Raspberry Pi) estimates in return the person's location with a small mean error of less than 10cm for the whole area. Results indicate that this approach holds strong potential for 6G use cases, particularly within the framework of Sensing as a Service.

## II. SYSTEM MODEL

In this paper we suggest the DNN-VLP indoor system, a reflection-based ΔRSS positioning method based on DNN using a monostatic device-free VLC system comprising a single LED transmitter, an array of ceiling PDs and a compute unit [3]. By ΔRSS we defined the difference of the RSS measured values due to obstacle, compared with expected – previously logged RSS values and indicates the position of obstacle in the room in a very fast and accurate way.

To evaluate the DNN-VLP system performance we assume a 5x5x3 $m^3$ room without furniture. Nine ceiling PD-based receivers (Rxs) are uniformly arranged in a 3x3 grid. LED transmitter (Tx) is centrally located on the ceiling and points directly downwards, with its normal vector aligned perpendicular to the floor. 8 PDs are facing downwards, tilted 10 degrees from their normal vector, and oriented towards the closest wall or corner of the room. One PD is co-located with the LED and faces directly downwards (Fig. 1). By tilting the 8 PDs and aligning them towards the room's edges, one ensures the detection of Non-Line of Sight (NLoS) light signals that have been reflected multiple times. DNN-VLP proposes to use the exact fluctuations on the NLoS light signals to provide significant information on the optical wireless channel. The compute unit is continuously monitoring the signals arriving from the PDs and when fluctuations are detected is inferencing the human target location through the already trained DNN model. The LED is utilized for both downlink transmission and lighting while PDs are used for receiving reflected transmitted power and positioning.

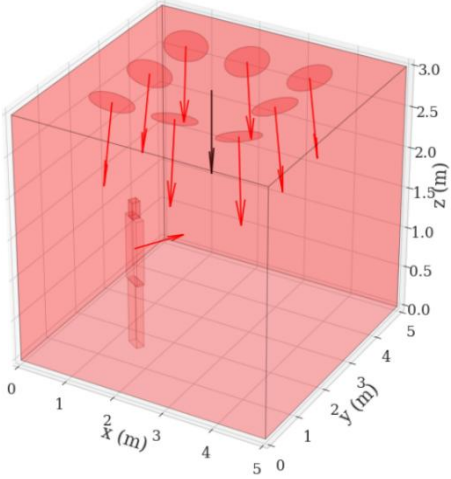

Fig. 1. DNN-VLP system setup. LED (black arrow) and PDs placed on the ceiling of room. Image produced with [7].

The VLC channel is modeled as a baseband linear system [8] expressed by

$$y(t) = R \cdot x(t) \otimes h(t) + n(t) \quad (1)$$

In this model, $x(t)$ denotes the optical intensity emitted by the LED, modulated by the input signal m(t). The output $y(t)$ corresponds to the photocurrent generated at PD. Here, R represents the responsivity of the PD, and $h(t)$ is the baseband Channel Impulse Response (CIR). The symbol $\otimes$ denotes the convolution operation. The term $n(t)$ accounts for additive white Gaussian noise (AWGN) which originates from the thermal noise, dark current, signal-induced shot noise, and background light-induced shot noise. CIR between a source and each receiver is modelled as an infinite sum over multiple reflections. Each reflection order is recursively computed using elemental objects, with the base case based on the DC channel gain $h^0(t)$, accounting for geometric and visibility factors between segments [9]. Here three reflections off the walls and floor are considered.

To model the system, room surfaces are considered completely diffusive and are segmented into smaller sections. Each segment is considered either a (light) source, a reflective surface, or a receiver. Sources are characterized by their position vector, normal vector, and Lambertian radiation mode number m. Similarly, the segments that act as receivers are defined by their position and normal vectors, the area of detection and their field of view (FoV). Reflectors are defined by their position, normal vectors and reflectance factor, while reflecting elements and LEDs are regarded as perfect Lambertian emitters ($m$=1).

The presence of a person near the LED-PD setup significantly affects the received signal strength (RSS), and makes possible the detection [10] and localization of targets—such as humans—by analyzing variations in reflection patterns. This technique, known as the relative RSS method or ΔRSS, relies on comparing two sets of measurements: one from an unoccupied room that serves as a reference, and another from the same environment when a person/target is present [3]. Building on this concept, the DNN-VLP approach proposed in this work operates in two stages: offline and online. During the offline phase, the RSS values recorded by the ceiling-mounted PDs in an empty room are initially calculated or measured.

Then, a grid of 49x49 points (2401 values), with a step length of 0.1 meters ranging from 0.1 to 4.9 in both the x and y axes is produced in the room and a human is placed successively at each grid point. A set of 9 RSS values detected by PDs are calculated or measured for each position. The ΔRSS difference, observed by each ceiling PD between the empty room and the room with the person at each grid location, constitutes the feature vector, i.e. the set of the ΔRSS values and respective coordinates of the person utilized to train the DNN-VLP ensemble. The outcome of the offline phase is a trained model.

During the online phase, real-time ΔRSS readings are fed into the trained model, which in turn estimates the person's location based on the learned patterns.

### III. DESIGN OF THE DNN-VLP ENSEMBLE MODEL

#### A. Selection of Model Architecture

Although both the REM prediction, as performed in previous work in [11], and the proposed VLP sensing experiments initially consider an empty room environment, the learning objectives and input data characteristics are fundamentally different. In the forward REM problem [8, 12, 13], RSS values are predicted directly from spatial coordinates under a single LED transmitter – single receiver, resulting in a smooth and well-conditioned regression task dominated by Lambertian propagation. In contrast, our inverse problem estimates/guesses target position using a multi-sensor RSS vector obtained from a ceiling photodetector array. This introduces strong cross-sensor correlations and nonlinear feature coupling, and requires from the model to infer obstacle position from spatial RSS fingerprints [14]. The inverse nature of the problem, combined with multi-input sensor fusion, increases significantly the complexity of the mapping function. These characteristics motivated our decision to implement an ensemble approach that combines three neural network types: MLPs [15], CNNs [16], and U-Nets [17], each contributing distinct capabilities to handle the problem's inherent ambiguities.

In this work, traditional fingerprinting methods such as K-Nearest Neighbors (KNN) [18] and tree-based regressors like ExtraTrees [19] is not included in the proposed ensemble for VLC-based ΔRSS localization. First, KNN is a non-parametric, memory-based approach that relies on distance-based searches over the entire training dataset. While effective for simple fingerprint matching, KNN is sensitive to noise and Euclidean distance metrics, which poorly captures the nonlinear and correlated structure of multi-sensor RSS vectors. Moreover, inference complexity scales linearly with the number of training samples, resulting in high latency and memory demands, which are incompatible with the real-time edge deployment objectives of the proposed system.

The ExtraTrees ensemble method is particularly effective for tabular inputs featuring weakly correlated or noisy variables, as its extreme randomization policy significantly reduces structural variance [19]. In the proposed system, the RSS inputs from multiple ceiling PDs form highly correlated, spatially structured vectors. ExtraTrees cannot exploit spatial dependencies nor perform hierarchical feature extraction, and its piecewise constant regression is less precise than the smooth nonlinear mappings learned by deep networks.

Finally, multi-architecture DNN-VLP ensembles inherently capture the roles traditionally performed by both KNN and

ExtraTrees. The CNN extracts spatial correlations among the sensor array, the MLP models global nonlinear relationships, and the U-Net provides dense reconstruction of spatial fingerprints for precise localization.

*B. Training Configuration*

The input data (49×49 grid) is partitioned into three subsets, with 60% used for model training, 20% reserved for validation, and the remaining 20% allocated for testing. Here the data is acquired through modelling of the VLC system however measured data could have been used. To ensure a fair comparison among the three architectures, data partitioning is not merely random.

Due to the RSS indoor distribution, we need to preserve a balanced spatial representation across all subsets, hence a stratified data splitting strategy is applied based on a 0.5m threshold from the room boundaries. This ensures that both interior and near-wall samples are included in each data set split.

We employed spatial k-fold cross-validation rather than conventional random partitioning with k-folding to perform weight optimization for the 9 individual instances comprising the ensemble. The key distinction lies in how we divided the data: K-Means clustering on position coordinates partitioned the room into geographically distinct regions before creating folds. This spatial separation between training and validation sets addresses a critical issue with indoor positioning data - nearby measurement points exhibit strong autocorrelation, so random splitting would allow models to essentially interpolate between neighboring training samples during validation rather than truly generalizing to new locations. Each fold designated one spatial cluster for validation while using the remaining clusters for training, then optimized ensemble weights by minimizing mean positioning error (MPE) on the held-out region. The final weights emerged from aggregating results across all folds using performance-weighted averaging. Folds achieving lower positioning errors received proportionally greater influence on the final weight values, effectively up-weighting the configurations that demonstrated better generalization to spatially distant regions. This approach yielded normalized weights that balance individual model contributions based on their validated localization accuracy across different areas of the room.

Larger k gives better granularity in spatial coverage so better detection of location-specific patterns. Large k also gives more reliable variance estimates, smaller validation sets per fold , without overfitting to specific regions. It's better for uneven spatial distributions where data cluster in certain areas, like near walls and corners in our VLP system, so more folds can isolate these better. Drawbacks of using many folds are longer computation time and risk of validating on regions that may not representative. To identify a suitable trade-off, we run the model for various k using our dataset. Based on this analysis, 3-fold spatial cross-validation provided the best balance between overall accuracy, robustness across the room, and computational efficiency, and was therefore adopted for the proposed DNN-VLP system.

To ensure fair comparison across architectures, we maintained consistent training parameters throughout: the Adam optimizer with a 0.001 learning rate, Mean Absolute Error (MAE) as loss function, batches of 32 samples, and 500 training epochs [11].

Finally, we trained three independent instances of each architecture using different random weight initializations to produce a total ensemble of 9 models. This repetition serves dual purposes, reduces sensitivity to initialization conditions and allows different instances to capture complementary aspects of the sensor-to-position relationship that a single model instance might miss.

*C. Model Configuration Details*

As we showed in our previous work [11] a shallow MLP (64×256) architecture is sufficient to approximate the forward prediction mapping but for inverse localization we employ a deeper architecture (64×256×64 ×256). This enables hierarchical feature extraction, latent representation compression, and refined position regression, and improves localization accuracy and robustness [14]. The intermediate bottleneck layer promotes feature compression and noise robustness, while the additional depth improves the model's capacity to capture reflection-induced RSS fingerprints. This design achieves improved localization accuracy and stability without excessive parameter growth, making it suitable for real-time device-free VLC sensing applications.

To evaluate the prediction capability of the model we assume (a) a human target to walk around a room in a trajectory of 25 consecutive steps and (b) the case that the target is placed in 100 random positions in the same room, and we let our models predict the position in both cases for the shallow and deep architecture. Both models predict well the position of the target with respect to the ground truth position, but the deeper architecture exhibits better accuracy for both (a) and (b).

This is illustrated in Fig. 2, where the deeper MLP configuration demonstrates a reduction in MPE of approximately 12–14% and an improvement in the 90th percentile error (P90) ranging from 7% to 32% relative to the baseline MLP (64×256).

The individual models of the DNN-VLP ensemble are MLP, CNN and U-Net architectures. MLP networks employ four fully-connected layers [64→256→64→256]. CNN architectures reshape the input vector into a 3×3 grid corresponding to the physical sensor layout, apply three convolutional layers (32, 64 and 128 filters with 3×3 kernels), and then connect to dense layers [64→256→64→256]. The U-Net architecture uses an encoder-decoder with skip connections (32→64→128 encoder filters, 128→64→32 decoder filters) to extract and reconstruct spatial RSS patterns, followed by the same 64→256→64→256 dense layers as the other two individual DNN models for position prediction.

## IV. RESULTS AND DISCUSSION

System parameters used for simulations are listed in Tab. I. OWCsim-Py [7], an open- source Python simulator for VLC [20, 21] is used to generate the training and inference datasets and Tensorflow to implement the DNN-VLP ensemble models. All calculations for timing recordings were executed on a workstation configured with a 4-core Intel(R) Xeon(R) E-2224 CPU @ 3.40GHz and 80GB of DDR4 RAM operating at 2666 MT/s.

TABLE I. MODELLING PARAMETERS OF DNN-VLP SYSTEM

| Parameter | Value |
|---|---|
| Room size | 5 x 5 x 3 $m^3$ |
| Location of LED ($T_x$) | (2.5, 2.5, 3) |
| Half Power Angle of LED | 60º |
| Transmitted power | 1000 mW |
| Area of PhotoDetectors (PDs) | $10^{-4}$ $m^2$ |
| Tilt angle of PDs | 10º |
| Responsivity of PDs | 1.0 |
| PDs Field of view (FOV) | 85º |
| Gain of optical filter | 1.0 |
| Refractive index of lens at the PDs | 1.5 |
| Reflectance factor of walls and floor | 0.8, 0.45 |
| Reflectance factor of human's hair, face, shirt | 0.6, 0.5, 0.3 |

We can include different combinations of individual DNN architectures to choose better accuracy or shorter model training time, or an intermediate state that balances the two model performance parameters. In Tab. II training time, mean position error and 90th percentile error are recorded for each ensemble All ensembles predict human's walking trajectory with MPE between 9.0 and 9.5 cm and the 100 human's random locations with MPE from 10.2 to 11.2 cm. The DNN-VLP ensemble that has by far the smallest training time (10% of all others) is the one that includes only the 3 instances of MLP 64x256x64x256, as expected due to small total number of trainable parameters (sum of all weights and biases across all layers) that is only 51k in respect to the 217k and 477k of the other two DNN models. Inference time is almost identical for all ensembles.

In Fig. 3 performance of DNN-VLP ensemble with MLPs is visualized for both cases. At the left the (Euclidean) positioning error (PE) heat map for the 100 uniformly selected random locations of the human target and at the right the prediction of a 25-step random walk trajectory (0.5 m per step) of the human target within the room.

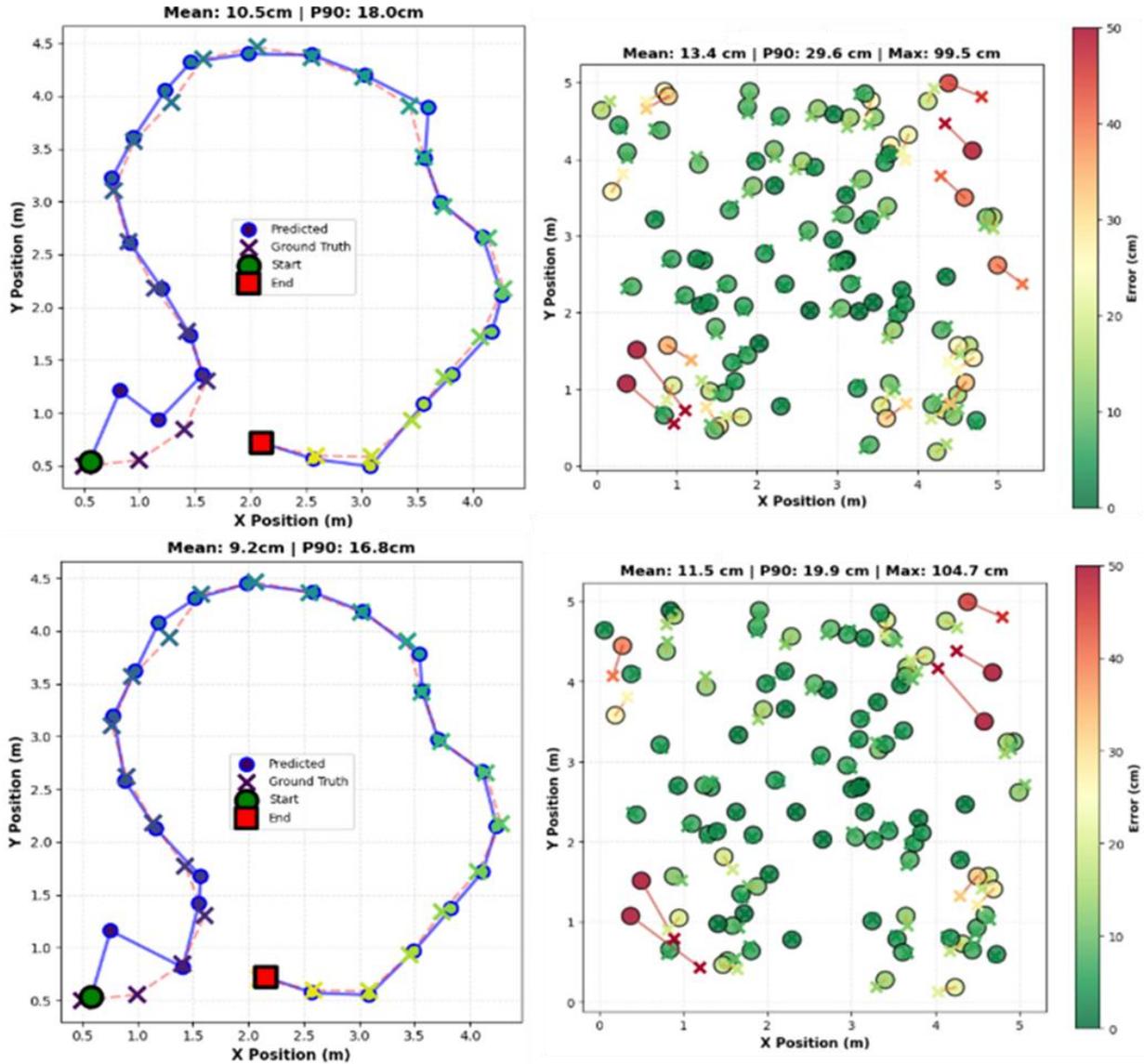

Fig. 2. Ground truth and MLP predictions of a 25-step random walk trajectory (0.5 m per step) of a human inside the room at the left, positioning Error (PE) values in cm at 100 uniformly selected random locations at the right. At the top are the MLP 64x256 prediction results and at the bottom the respective predictions for MLP 64x356x64x256.

TABLE II. PERFORMANCE OF DNN-VLP ENSEMBLES

| DNN-VLP Ensemble of three instances per architecture | Training time (min) | Walking trajectory | | 100 random locations | |
|---|---|---|---|---|---|
| | | *MPE (cm)* | *P90 (cm)* | *MPE (cm)* | *P90 (cm)* |
| MLP | 1.5 | 9.39 | 12.44 | 11.21 | 18.60 |
| MLP and U-Net | 12.8 | 9.35 | 16.87 | 10.19 | 16.17 |
| MLP, CNN and U-Net | 16.9 | 9.07 | 11.79 | 10.33 | 19.57 |

In our analysis we utilize the mean distance Error as a metric. However, for training the DNN-VLP model, we opt for MAE, which represents the Manhattan distance between the predicted and actual location points, as loss function. MAE is the average of the absolute differences between the predicted and actual values, while MSE is the sum of squared Euclidean distances. Fig. 3 shows that when the human target is positioned close to the walls of the room the positioning errors are larger than the ones found for locations in the middle of the room under the LED. With MAE as loss function, the model is less sensitive to outliers and performs better when large errors are rare but extreme, while MSE pulls predictions more strongly toward outliers because errors are squared.

In developing ML based VLP systems, the shift from single-model architectures to DNN ensembles allows for more adaptable systems. By decoupling the offline training phase from the real-time inference phase, we can curate a "library" of specialized ensembles. The versatility of an ensemble system serves vastly different operational requirements within the same physical infrastructure. For example, if autonomous warehouse robots are located a high-accuracy ensemble is needed as robots require sub-centimeter precision to navigate narrow aisles and interact with shelving units. Training time here is secondary to minimizing collision risks. For occupancy-based lighting control a fast-training ensemble is more appropriate as presence sensing is the priority. For the case of interactive museum guides a balanced/intermediate ensemble gives enough accuracy to trigger audio for a specific exhibit (decimeter level) while maintaining low computational overhead for mobile device integration.

DNN-VLP is a single target device free monostatic system comprising a single LED, a configurable set of ceiling mounted PDs and a single board computer (p. ex. Raspberry Pi). In comparison to other similar systems [3], it achieves comparable positioning outcomes in a cost-effective manner. Further optimization with respect to number of LEDs, number of PDs, finer human positioning during offline training is expected to exhibit more accurate positioning prediction with the same DNN system and compute node complexity.

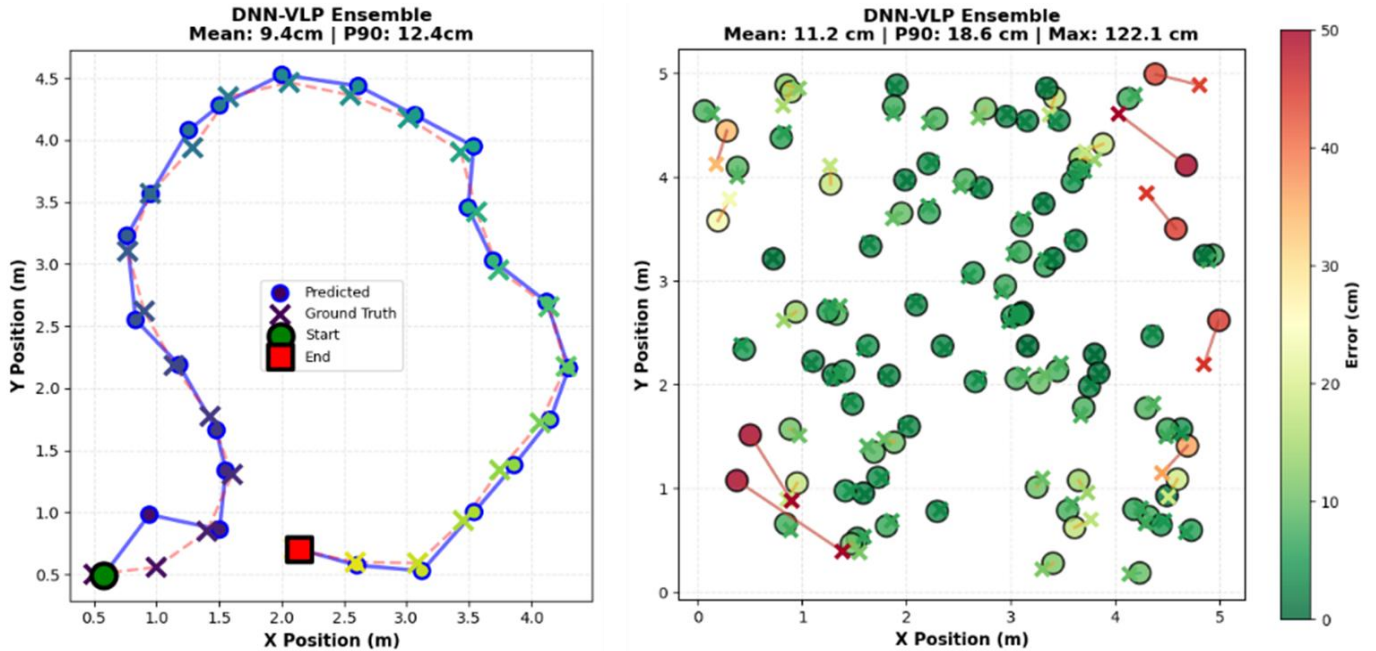

Fig. 3. Ground truth and DNN-VLP Ensemble of MLPs predictions of a 25-step random walk trajectory (0.5 m per step) of a human target inside the room at the left, positioning Error (PE) values in cm at 100 uniformly selected random locations at the right.

## V. Conclusions

In this paper, we developed a device-free indoor positioning and sensing system using VLC, employing multi-architecture DNN-VLP ensemble models. The proposed DNN-VLP system demonstrates positioning with errors under 10 cm within the LED's cell and offers ultra-fast prediction performance. In a dynamic environment where moving targets create complex shadowing, the DNN-VLP system can adapt by selecting an ensemble tailored to the task. It prioritizes high-precision models for critical object tracking, while switching to lightweight models for general tasks. By conceding a negligible 1 cm in accuracy, the system slashes training time and energy consumption, enabling the positioning engine to update near-instantly to possible room layout changes.

## Acknowledgements

H. S. and C.T.P would like to acknowledge Empeirikion Foundation and European Commission, Horizon Europe, Grant Agreement No 101087257, METACITIES Exhellence Hub for funding this work.